# Physical collisions of moonlets and clumps with the Saturn's F-ring core

**Sébastien CHARNOZ**
*Equipe AIM, Université Paris Diderot/CEA/CNRS*
*CEA/SAp, Centre de l'Orme Les Merisiers*
*91191 Gif-Sur-Yvette Cedex*
*France*

Pages = 17

Figures = 6

Tables= 1

* : To whom correspondence should be addressed  : charnoz@cea.fr

**Abstract:** Since 2004, observations of Saturn's F ring have revealed that the ring's core is surrounded by structures with radial scales of hundreds of kilometers, called "spirals" and "jets". Gravitational scattering by nearby moons was suggested as a potential production mechanism; however, it remained doubtful because a population of Prometheus-mass moons is needed and, obviously, such a population does not exist in the F ring region. We investigate here another mechanism: dissipative physical collisions of kilometer-size moonlets (or clumps) with the F-ring core. We show that it is a viable and efficient mechanism for producing spirals and jets, provided that massive moonlets are embedded in the F-ring core and that they are impacted by loose clumps orbiting in the F ring region, which could be consistent with recent data from ISS, VIMS and UVIS. We show also that coefficients of restitution as low as ~0.1 are needed to reproduce the radial extent of spirals and jets, suggesting that collisions are very dissipative in the F ring region. In conclusion, spirals and jets would be the direct manifestation the ongoing collisional activity of the F ring region.

# 1. INTRODUCTION

Saturn's F ring, located about 3400 km beyond the edge of Saturn's A ring, is a dusty ringlet well known for exhibiting a large variety of transient structures (see, e.g., Smith et al., 1981, Murray et al.,1997, Poulet et al.,2000, Showalter 2004, Murray et al. 2005a, 2005b, 2008). Whereas the close presence of Prometheus seems to explain some periodic features (Showalter and Burns 1982, Murray et al., 2005), other large scale-radial structures are present around the F-ring core: "spirals" (Charnoz et al. 2005, see Figure 1) and the related phenomenon of "jets" (Murray et al., 2008, Figure 1 and Figure 2) originating from the core. They both have been suggested to be the results of collisions of nearby moonlets with the central region of the F ring. The spiral (Figure 1) appears as ringlets parallel to the F-ring core. They were believed previously to be arcs of material parallel to the core, and were first called "strands" (Murray et al., 1997), but in fact the Cassini spacecraft has revealed that they are connected to form a single spiral structure as a result of Keplerian shear (Charnoz et al. 2005). At the intersection of these spirals with the F-ring core, smaller chaotic structures of scattered material called "jets" are observed (Charnoz et al. 2005, Murray et al. 2008). All these structures are bright at high phase angles so may be mainly composed of dust (Showalter et al., 1992, Charnoz et al., 2005).

What is their origin? Charnoz et al. (2005) originally proposed that spirals are composed of particles gravitationally scattered from the F-ring core by a Prometheus sized satellite, admitting however that such a large satellite should have already been identified. At the same epoch, several bodies, that could be either moonlets or clumps, were discovered orbiting in the F-ring region. Among them S/2004 S6 was especially interesting because its orbit intersects the F-ring's core (Porco et al., 2005). This is why Charnoz et al. (2005) suggested instead that physical collisions of F-ring material with a kilometer-sized moonlet could be a possible mechanism for producing spirals, without proving it analytically or numerically. This mechanism is also invoked in Murray et al. (2008) to explain the origin of jets, but again, they do not really investigate the mechanism of physical collisions. This simple idea is supported by the fact that a population of nearby moonlets and clumps, suspected for long (Cuzzi & Burns 1988), has been recently detected in some high resolution Cassini images (Murray et al. 2005b, Porco et al. 2005). Stellar occultations tracked by the UVIS instrument onboard Cassini allowed the F ring's densest regions to be probed at high resolution and revealed the presence of 500m to 1.5km bodies embedded in the core (Esposito et al. 2008). All these new moonlets and clumps have eccentricities of a few $10^{-3}$ (Murray et al. 2005a, Spitale et al. 2005, N.J. Cooper, private communication 2005). Their elongated appearance (Porco et al., 2005) suggests that they may be loose aggregates of material. In conclusion two populations of moonlets may exist: one exterior to the ring and another embedded in the F ring's core.

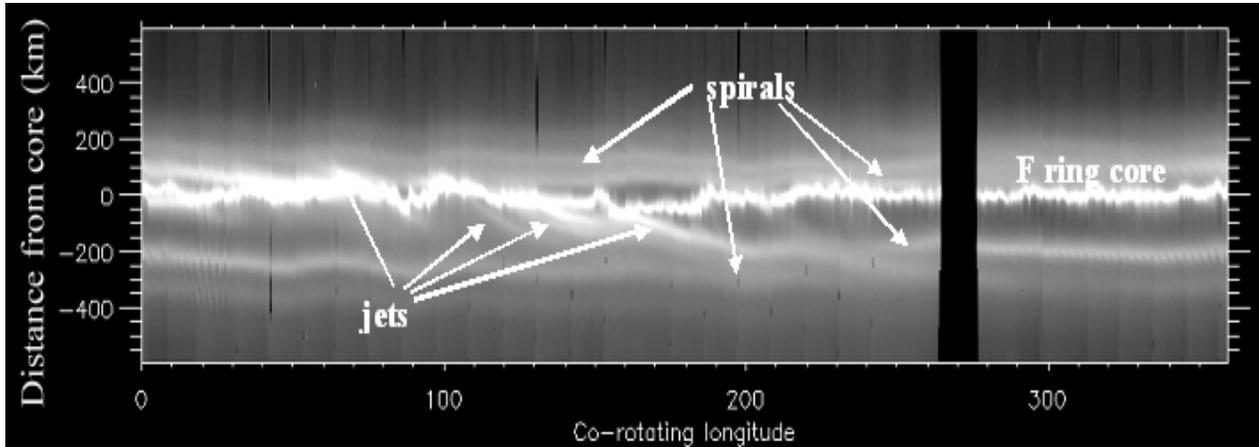

*Figure 1*: (up) Mosaic of the F ring taken in April 2005 built from a fixed observation point in the Saturn's inertial frame (49° longitude). From Charnoz et al. (2005). The resolution is 65 km per pixel. Images have been re-projected in a rotating frame moving at the F ring orbital speed. Spirals, originating from jets, are clearly visible. Jets at higher resolution are visible in Fig.1.

Among the common characteristics of the spirals and jets, we note: (i) their wide radial amplitude (ii) their relative radial locations to the F-ring core seems to be an effect of semi-major-axis difference rather than an eccentricity effect (Murray et al., 2008). These elements suggest that the orbital elements of material in jets and spirals may have eccentricities and longitudes of pericenters close to the F ring's orbit (Bosh et al., 2002), with a large dispersion in semi-major axes (± a few 100 km), explaining both the radial location of material and the rapid shearing of jets to form spirals. These elements may constrain the dynamical origin of these structures.

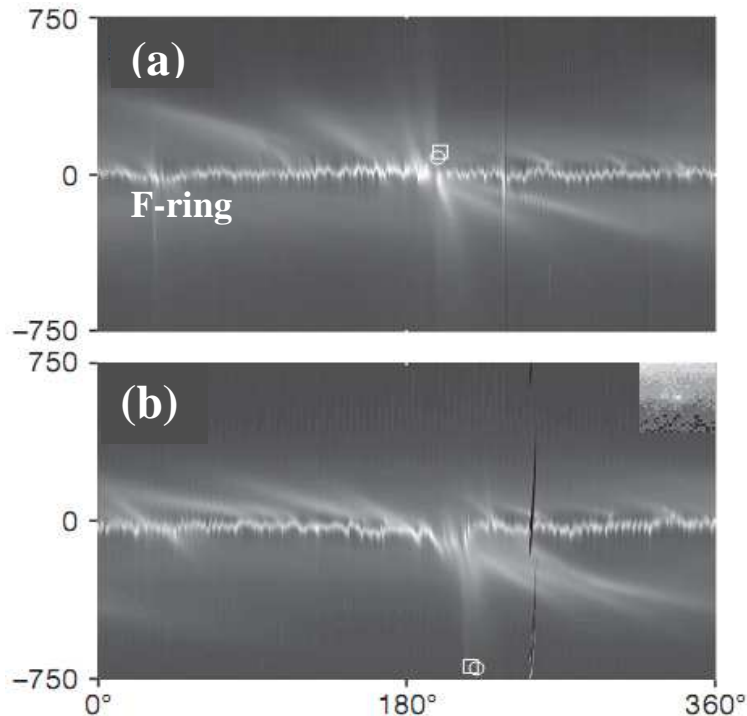

*Figure 2: Cassini observation of the F ring, in a longitude/radius frame. On the vertical scale, units are in kilometres from the F ring core. From Murray et al., 2008. (a) March 2007 (b) April 2007. The circle designates the predicted location of S2004/26 (Murray et al., 2008) and the square the predicted location of a new object reported in Murray et al. 2008.*

In the present study we quantify the effect of a physical collision of massive kilometre-sized moonlets on the orbit of massless particles contained in ringlets, to see if this could be a viable mechanism to produce jets and spirals around the F-ring core. In the next section we first give analytical insights into the dynamics of the encounter. In the third section we use numerical simulations to follow both the short-term and long-term evolution of the structures. In section 4 we investigate the case of a cohesionless clumps impacting cohesive and massive bodies embedded in the F ring. These results are discussed in the last section.

## 2. A simple model

We assume here that a moonlet with mass $m_s$ encounters a cloud of particles having individual masses $m_p$ with $m_p \ll m_s$. In addition, the moonlet has semi-major axis $a_s$ and eccentricity $e_s$ whereas those of the particles are $a_p$ and $e_p$, respectively. Since both are located in the region of the F ring $a_s \sim a_p$ and eccentricities are of the same order of magnitude ($e_s \sim e_p$). We now estimate the distance over which the particles are scattered during a close interaction, either a gravitational encounter or a physical collision, in order to properly compare both mechanisms. In the case of gravitational scattering, the magnitude of the orbital velocity perturbation, $\Delta V_g$, is of the order of the escape velocity of the satellite (assuming that the distance at close approach is comparable to the satellite's physical radius).

$$\Delta V_g \approx \sqrt{\frac{2Gm_s}{r}} \qquad \text{Eq.1}$$

To scatter material over $\Delta r \sim 300$ km, $\Delta V_g$ must be at least of the order of the difference of orbital velocity across this distance, which is about $\Delta r \times (GM_s/a_s^3)^{1/2} \sim 17.5$ m/s, where G is the universal gravity constant, and $M_s$ is Saturn's mass. Solving Equation 1 for $m_s$, we find that $m_s$ corresponds to body of radius 33 km (assuming Prometheus' density) about the size of the shepherd moons Pandora or Prometheus. If such a body existed in the F ring region, it would have been already detected, furthermore Pandora and Prometheus do not approach the F-ring core close enough to produce spirals and jets (Charnoz et al. 2005, Murray et al. 2005b, 2008). Consequently nearby moonlets of a few kilometers only (as observed) cannot produce such wide structures via simple gravitational scattering. They can only perturb the material over short distances (~10 km) due to their gravitational effect (Charnoz et al. 2005, Murray et al. 2008).

We now consider physical collisions. The usual inelastic rebound formalism (see, e.g., Richardson 1994) implies that after an impact, the relative velocities are such that:

$$\mathbf{U'} = -\varepsilon_n \mathbf{U_n} + \varepsilon_t \mathbf{U_t} \qquad \text{Eq.2}$$

where **U'** is the post-impact relative velocity, **U$_n$** is the pre-impact relative velocity normal to the surface of both bodies, **U$_t$** is the pre-impact relative velocity tangential to the body's surface, $\varepsilon_r$ and $\varepsilon_t$ the radial and tangential coefficients of restitution (between 0 and 1). We call **V$_p$** and **V$_p$'** the pre- and post-velocities of a particle in the inertial frame and $\mathbf{V_s}$ and $\mathbf{V_s}$' the pre- and post-impact velocities of the satellite in the inertial frame. By definition $\mathbf{U_n}=\mathbf{R}(\mathbf{V_s}-\mathbf{V_p})\cdot\mathbf{R}$ with $\mathbf{R}$ standing for the unit vector joining the particle to the satellite's center, we get also $\mathbf{U_t}=(\mathbf{V_s}-\mathbf{V_p})-\mathbf{U_n}$ (in which we have neglected the body's spin for the computation of the tangential relative-velocities). After solving for conservation of linear and angular momentum, the post-impact velocity of a particle is given by (eq. 15 in Richardson 1994):

$$\mathbf{V_p'} = \mathbf{V_p} + (1+\varepsilon_n)\mathbf{U_n} + \beta(1-\varepsilon_t)\mathbf{U_t} \qquad \text{Eq. 3}$$

where β is a parameter between 0 and 1 depending on the moment of inertia of the colliding bodies. For uniform spinning spheres β=2/7, whereas for point-mass or non-rotating bodies β=1 (Richardson 1994). After replacing $\mathbf{U_t}$ in the above expression, we get the post-impact velocity of the particle as a function of $\mathbf{V_s}$, $\mathbf{V_p}$ and $\mathbf{U_n}$

$$\mathbf{V_p'} = \mathbf{V}_p + (1+\varepsilon_n - \beta + \beta\varepsilon_t)\mathbf{U_n} + \beta(1-\varepsilon_t)(\mathbf{V}_s - \mathbf{V}_p) \qquad \text{Eq.4}$$

The satellite's velocity $\mathbf{V'_s}$ remains unchanged (i.e., $\mathbf{V'_s}=\mathbf{V_s}$) because $m_s>>m_p$. Interestingly for β≠1 we note that Eq.4 predicts that even for a fully dissipative collision ($\varepsilon_n=\varepsilon_t=0$) the two objects cannot physically stick together. Indeed for $\varepsilon_n=\varepsilon_t=0$ Equation 4 implies that the relative velocities between the two center-of-masses after the impact, $\mathbf{V'_p}-\mathbf{V_s}$, is $(\mathbf{V_p}-\mathbf{V_s}+\mathbf{U_n})\times(1-\beta)$ which never equals 0 for β≠1 (unless $\mathbf{V_p}-\mathbf{V_s}=\mathbf{U_n}$, which means a perfectly head-on collision, but that virtually never happens). Why does the post-impact relative velocity between the center-of-masses never reaches 0, even for a perfectly dissipative collision? Equation 4 allows for the two spheres to roll on each other with some friction (quantified by $\varepsilon_t$) so a fraction of the relative kinetic energy is stored in the spin and redistributed to the particles' center-of-mass after the impact (Richardson 1994). It results in an "effective" tangential restitution coefficient that cannot be smaller than (1-β), =5/7 for spheres, which is very high (only 30% of the relative tangential velocity disappears at most for spherical bodies !). It is a limitation of the smooth spheres model. Whereas this behaviour may be realistic for smooth spheres of similar sizes, it is very doubtful for small particles impacting a satellite's surface at high velocity (~30 m/s). Complex surface topology, multiple rebounds on a chaotic surface, the presence of a regolith layer, all these effects combine to make such a rebound highly dissipative and efficient to damp the spin of the incoming particle. A dust particle impacting a satellite on a grazing trajectory may suffer a reduction of relative velocity by a large factor, *a priori*, much greater than 30% (for β=2/7) as predicted by Equation 4. In conclusion, we think that it would be more realistic for our case to set β=1 since we are dealing with highly dissipative collisions (dust impacting on a satellite's surface). It has also the advantage of treating the normal and tangential velocities in symmetrical ways, which is appropriate since on a small satellite with a complex surface, the local direction normal to the surface can be random. When setting β=1, Eq. 4 reduces to

$$\mathbf{V_p'} = \mathbf{V_s} + (\varepsilon_n + \varepsilon_t)\mathbf{U_n} - \varepsilon_t(\mathbf{V_s} - \mathbf{V_p}) \qquad \text{Eq.5}$$

If $\varepsilon_n$ and $\varepsilon_t$ are set to 0 (perfect sticking), the particle's post-collision velocity will be exactly the same as the satellite's velocity. It means that when a cloud of massless particles encounter a satellite, in the limit $\varepsilon_n=\varepsilon_t=0$, those particles that bounce on the satellite's surface are put on an orbit identical to the satellite's orbit due to the conservation of momentum. When $\varepsilon_n$ and $\varepsilon_t$ are larger than 0, the particles will spread around the satellite's orbit, so their semi-major axes will be on average the satellite's semi-major axis $a_s$, with a dispersion $\Delta a$ around this central value. Now we estimate the typical semi-major axes dispersion $\Delta a$, to quantify the distance to which the material spreads after any collision. The post-impact orbital energy per mass-unit is given by the following equation (in which we only kept first order terms in $\varepsilon_n$ and $\varepsilon_t$) :

$$E = -\frac{GM_s}{r_p} + \frac{1}{2}V'^2_p \qquad \text{Eq.6}$$

$$E = -\frac{GM_s}{r_p} + \frac{1}{2}\left(V_s^2 + 2\mathbf{V_s} \cdot (\varepsilon_n + \varepsilon_t)\mathbf{U_n} - 2\mathbf{V_s} \cdot (\mathbf{V_s} - \mathbf{V_p})\varepsilon_t + o(\varepsilon)\right) \qquad \text{Eq.7}$$

$$E = E_s + \Delta E \qquad \text{Eq.8}$$

with $E_s$=-$GM_s/r_p$ +1/2$V_s^2$ being the Satellite's specific orbital energy (since $r_p$, the radial distance to Saturn is the same for the satellite and particles at the moment of impact); and the remaining terms are defined as the spread in specific orbital-energy $\Delta E$. Due to random orientations of vector $\mathbf{V_s}$ with vectors $\mathbf{U_n}$ and ($\mathbf{V_s}$-$\mathbf{V_p}$), for a cloud of particles the dot-products may vary between ±1 so that $\Delta E$ will always have values distributed between ±[$V_s(\varepsilon_n+\varepsilon_t)U_n+2V_s(V_s-V_p) \varepsilon_t$]. This will induce, in turn, a spread in semi-major axes given by (Murray and Dermott 1999):

$$\Delta E = \frac{GM_s}{2a^2}\Delta a \qquad \text{Eq.9}$$

Noting that the magnitude of $\mathbf{U_n}$ and $\mathbf{V_s}$-$\mathbf{V_p}$ are both of the order of the relative velocity between the two bodies $\sim eV_k$ (with *e* standing for the average eccentricity and $V_k$ for the local keplerian velocity), that $V_k \sim (GM_s/a)^{1/2}$ , and that $V_s=V_k$ we get

$$\Delta a \approx \pm 6ae\varepsilon \qquad \text{Eq.10}$$

where we have assumed $\varepsilon_n \sim \varepsilon_t \sim \varepsilon$ for simplicity. Eq. 10 gives the full range of accessible semi-major axes, but does not provide the exact distribution inside these ranges. This distribution depends on the specific geometry of impact and on the shape of the satellite, but Eq.10 provides an estimate of the full extent of the radial spreading, which is enough for our qualitative study. In the F-ring region, typical eccentricities are about 0.002 (making encounter velocities about 25 m/s), so in order to produce a structure extending over 300 km, a coefficient of restitution of only 0.18 is sufficient, meaning that 97% (1-0.18$^2$) of the relative impact energy is absorbed during the impact. We note that ±300km is the upper bound of the spiral's radial extent, and in average the material is gathered between ±200 km around the core (see Figure 1 and Figure 2), so that a coefficient of restitution as small as 0.1 may be enough to explain the majority of the radial extent. So physical collisions of dust on a satellite's surface is a very promising mechanism to produce a widely extended radial structure, in particular because such a process has almost no dependence on the satellite's mass, provided it is much higher than a dust particle's mass. This absence of mass dependence allows a small (radius ~1km) satellite to scatter dust over hundreds of km, whereas it is impossible for pure gravitational scattering (see above and in Charnoz et al. 2005). To go beyond these simple considerations, we now turn to a numerical simulation, to explore in more detail the dynamics of such an interaction, and to examine the long-term evolution of the structures created.

## 3. Numerical Simulation of a solid moonlet encountering a massless F ring

To simulate the physical encounter of massless particles with a massive moonlet crossing the F-ring core, the orbits of 10,000 test particles were integrated, as well as the orbit of a single massive body (the moonlet), using the numerical procedure described in Charnoz et al. (2000). Test particles experience gravitational interaction and dissipative collisions with the moonlet. Initially the test particles are gathered into a small segment with a ±5 km spread around the F ring's orbit (a=140223,e=0.00254, i=0.0065°, Bosh et al. 2002). The moonlet's orbit is set to one possible orbit of S/2004 S6, and is considered as typical of other moonlets discovered in this region (N.J. Cooper, J.N. Spitale, private communications, 2007) with a~140140 km, e~0.002, i~0.002° (as representative values). The orbital orientations (longitude of perihelion and of ascending node) were chosen so that the moonlet penetrates into the swarm of test particles on the F ring's orbit. Different images showing the impact are displayed in Figure 3. Just after the moonlet penetrates the F ring (Fig.3a) test particles that suffered a rebound on the moonlet's surface follow the moonlet's center of mass (by simple conservation of momentum, since test particles have null mass whereas the moonlet has a finite mass), and organize as a spherical shell around the moonlet reflecting the moonlet shape, just like a specular reflection at reduced speed. After some time, this structure spreads longitudinally due to keplerian shear and creates an elongated structure, with a "banana" shape centered around the moonlet (Fig.3b, Fig.3c), which may be similar to some of the elongated clumps observed in the F ring (Porco et al. 2005).

The orbital elements of those particles that bounced off the moonlet's surface are displayed in Fig.4, assuming tangential and radial restitution coefficients about 0.05. In the semi-major-axis / eccentricity space, particles that collided with the moonlet are grouped around the moonlet's orbital elements and show a wide spread in semi-major axis with variations over hundreds of kilometres. The dispersion of eccentricity is about $5 \times 10^{-4}$, resulting in ~70 km of radial excursion. So the radial excursion of material scattered from the core is mainly an effect of semi-major axis variation rather than eccentricity. Rapidly (after ~1 orbit) the moonlet gravitationally interacts with the scattered particles and creates a central void around itself of a few Hill radii, clearly visible in the Fig.3.c. The total radial excursion, around the satellite's orbit, as a function of the coefficient of restitution was computed from different numerical experiments. They are reported in Table 1 and in Fig.5 and are in agreement with the analytical estimate given in Equation 10 ( as shown in Fig.5, results of table 1 are reasonably reproduced by Eq.10 using e~0.002 and a~140200 km). Numerical simulations suggest that a coefficient of restitution ~ $5 \times 10^{-2}$ is needed to reproduce the radial extension of observed structures in the F ring region (Fig. 1) in reasonable agreement with our previous estimate about 0.1. So, only 0.2% of the relative impact kinetic energy is conserved during an impact ($0.05^2$). Such very low coefficients of restitution are expected in high velocity impact as suggested in laboratory experiments (Bridges et al. 1984, Supulver et al., 1995). On long-term evolution, as the scattered material shears due to its initial spread in semi-major axis, a spiral is created (as described in Charnoz et al. 2005). The number of turns in the spiral structure increases linearly with time due to the keplerian spreading of particles. But as a direct consequence of conservation of momentum, the ejecta are spread symmetrically around the system's center-of-mass' orbit, which is the satellite's orbit itself here, *rather than around the F-ring core's orbit* (which is assumed to contain no mass in the present simulations). This is in contradiction with observations (Fig. 1), in which spirals are clearly parallel to the F-ring core (meaning that the pericenters' longitudes have similar values), as well as jets that seem to originate also from the core itself (Fig. 2). To solve this discrepancy, the role of the moonlet and particles are reversed in the next section.

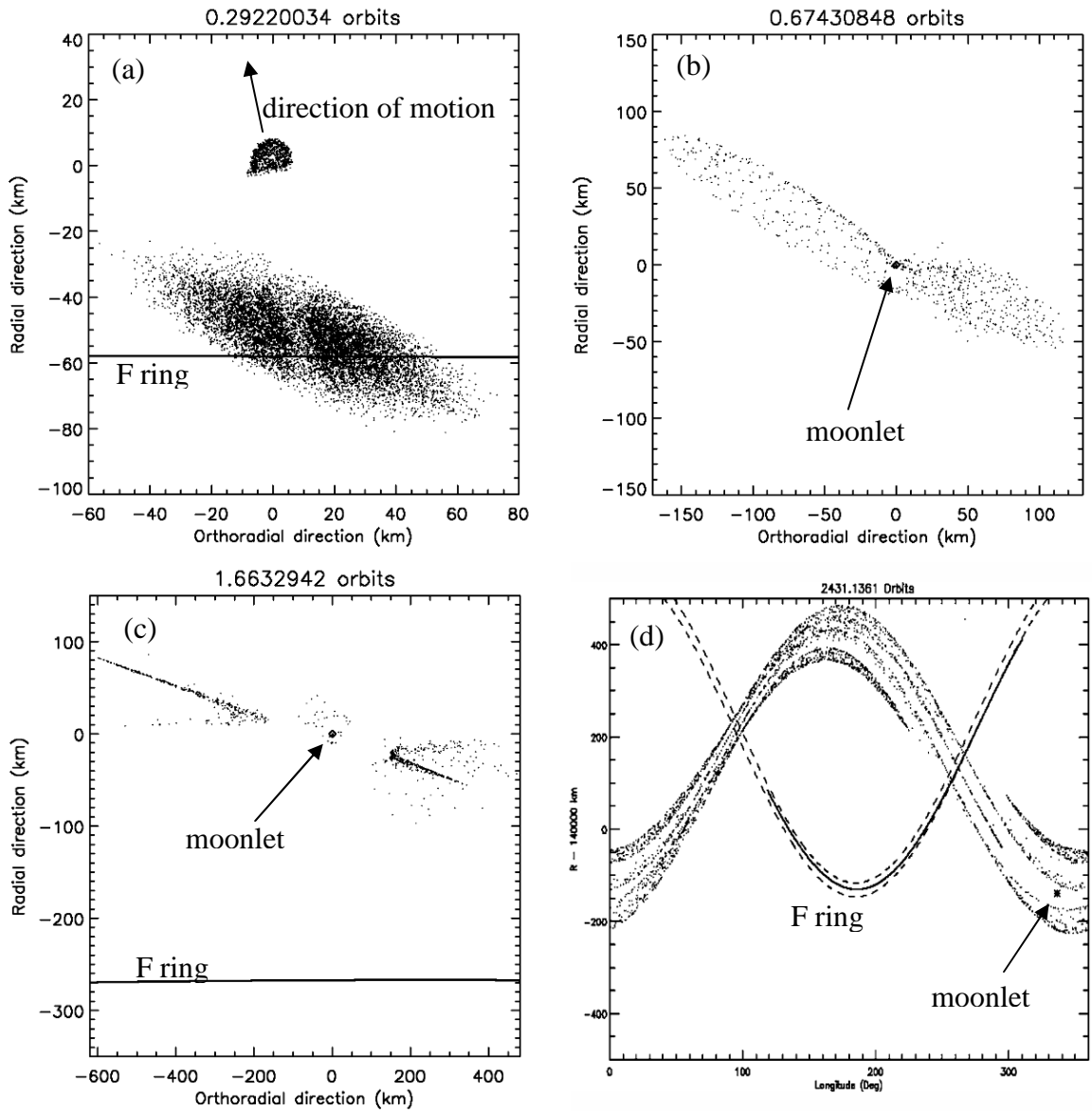

*Figure 3*: *Snapshot of the test particles just after the moonlet impacts the particles gathered in the F ring. Test particles are symbolized by dots and the moonlet by a diamond. Positions are given in Km relative to the moonlet's center. The Y axis is the radial direction, the X axis is the ortho-radial direction. (a) ~0.3 orbit after impact: the moonlet digs a hole in the F ring, (b) 0.8 orbit after impact. (c) after 1.6 orbit: a central hole is visible in the clump around the moonlet, due to gravitational scattering and re-accretion,(d) locations of test particles after 2430 orbits, in a longitude-radius reference frame: a spiral is created. The two dashed lines represent the location of the F ring.*

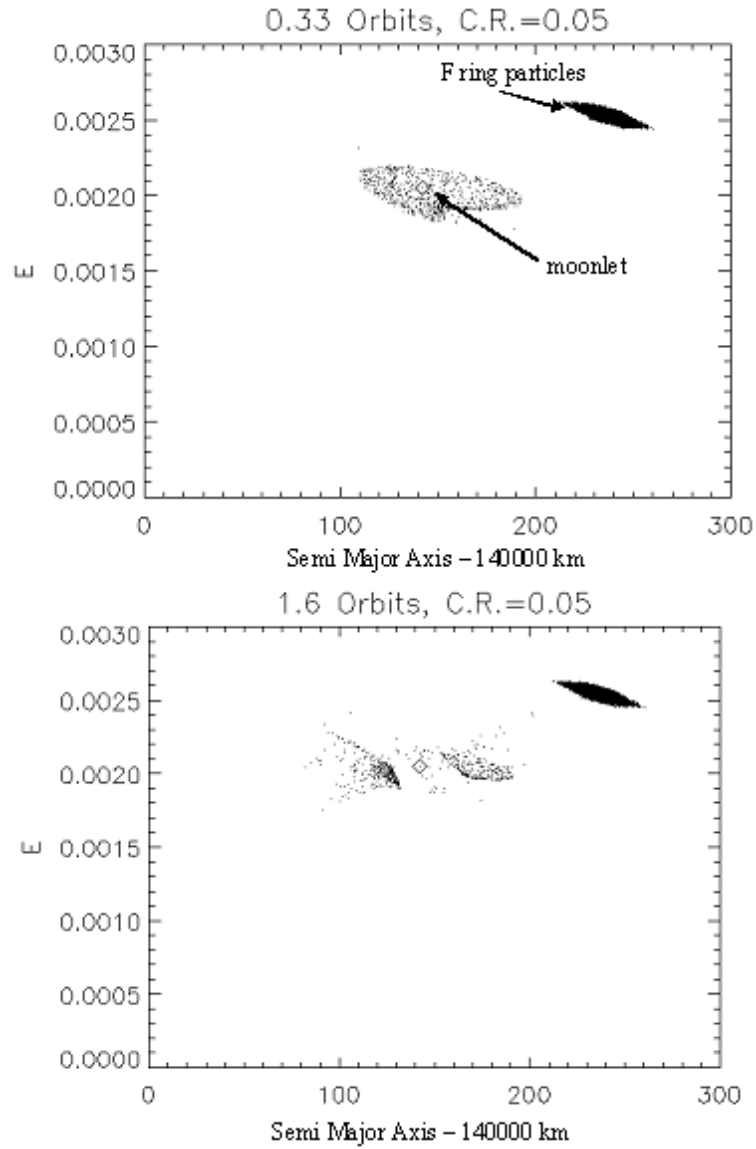

*Figure 4*: *Orbital elements (semi-major axis versus eccentricity) of test particles (dots) and moonlet (diamond) just after the impact (top) and after 1.6 orbits (bottom). In this simulation radial and tangential coefficients of restitution are set to 0.05.*

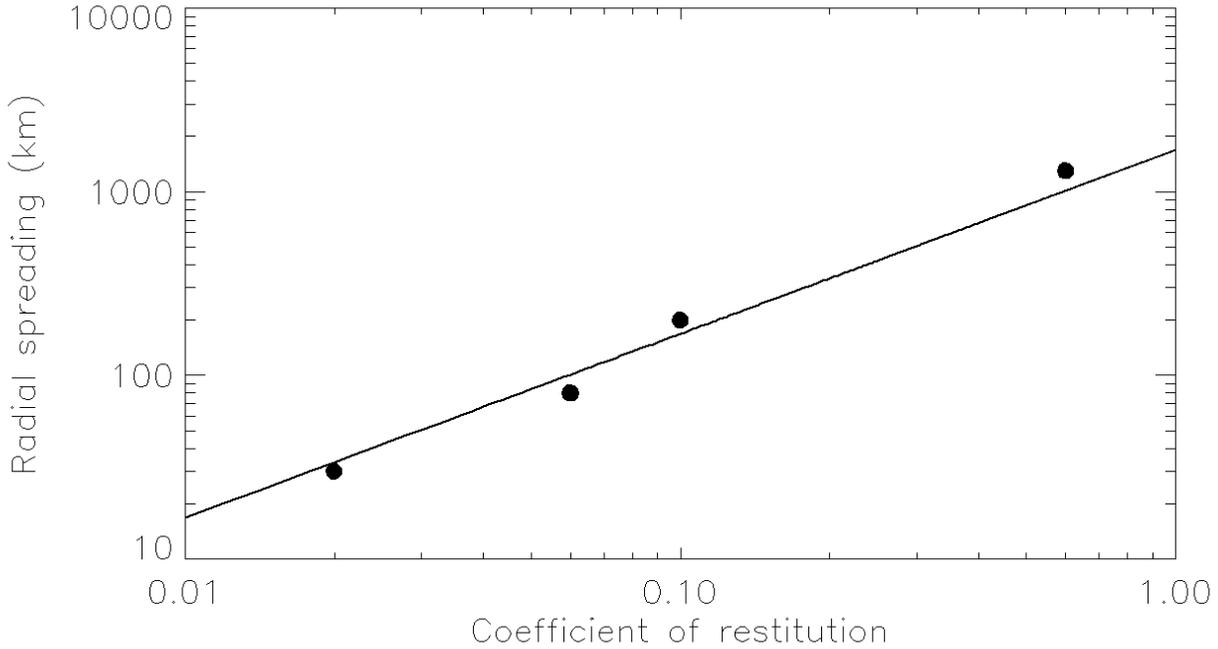

*Figure 5*: *Radial spreading of dusty material around the F ring core after a collision with a moonlet as a function of the coefficient of restitution (same values for the normal and tangential coefficients). Black dots: data from numerical simulations (Table 1). Solid line: Equation 10 using a=140200km and e=0.002.*

### 4. A massless clump encountering a massive moonlet inside the F ring

Several bodies observed around the F ring are elongated (Porco et al 2005, Murray et al., 2008) and could be clumps of material rather than solid and spherical bodies. They could be also a combination of the two: a central core surrounded by a cloud of particles. So we explored another scenario in which the body impacting the F ring is a very loose aggregate of particles, called a "clump", colliding with a solid moonlet embedded inside the F-ring core. This idea is motivated by two arguments (i) the F-ring core ring is now known to shelter a small population of moonlets (Cuzzi and Burns 1988, Esposito et al. 2008, Murray et al. 2008) and (ii) the spiral winds around the system's center-of-mass (see previous section) and so, we need the F-ring core to be itself the center-of-mass. It is why we assume now that there is a solid and massive moonlet embedded in the F-ring core. We call it the "F-ring moonlet". To simulate the impact of the clump with the F-ring moonlet, we gathered 10000 test particles in a sphere of 5 km radius, on an S6 like orbit (see previous section) to simulate the clump. All test particles are insensitive to each other, so the clump is completely cohesion-less. The F-ring moonlet is simulated by a 5 km radius massive body put on the F ring's orbit (and corresponding mass for a density of 0.9 g/cm$^3$). The clump impacts the F-ring moonlet and is scattered appart. Results are shown in Fig.6. Just after the impact, a spherical cloud of test particles forms around the F ring moonlet.

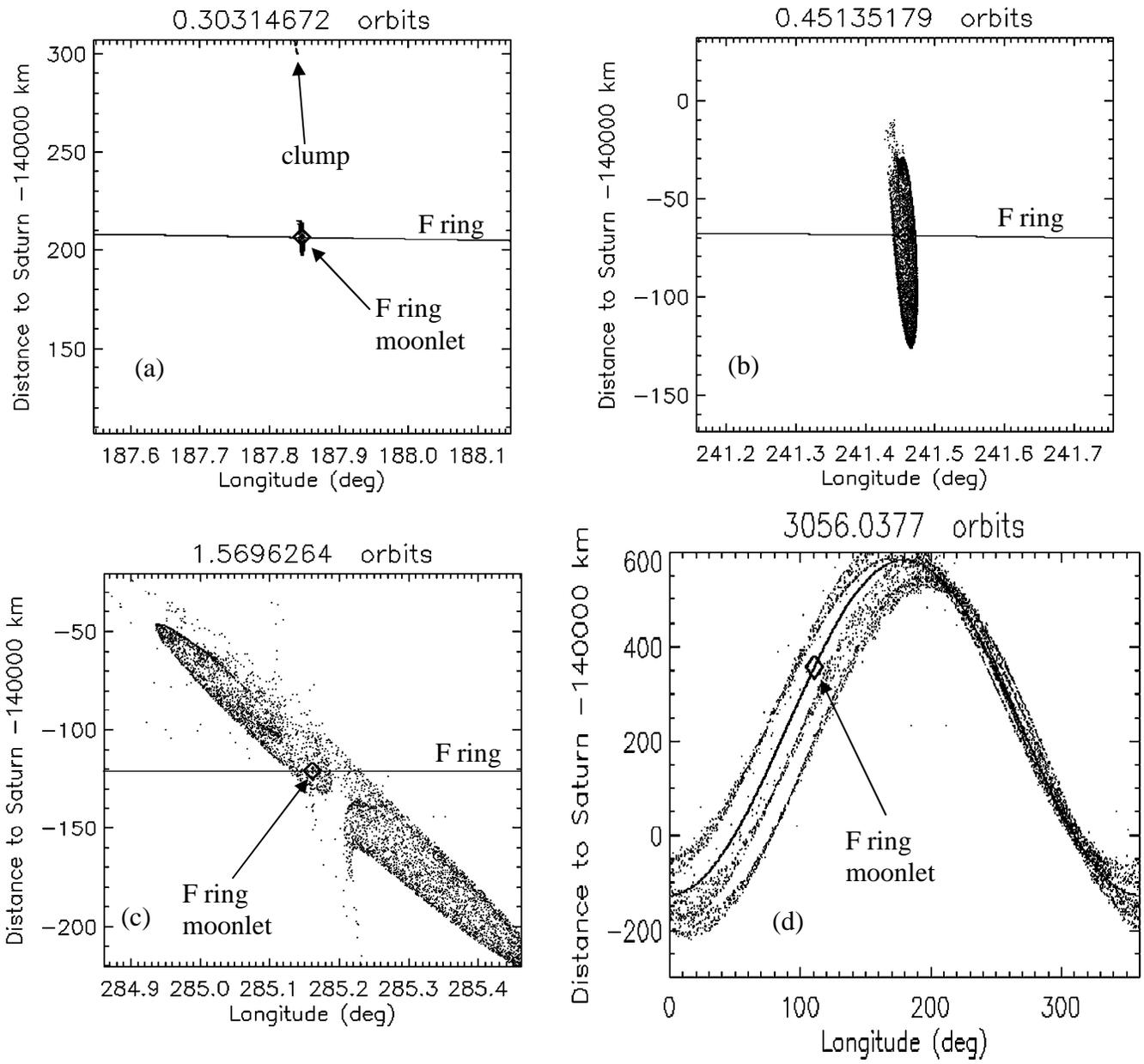

*Figure 6: Time evolution of particles in the case of a loose clump (made of test particles, dots) impacting a massive satellite (diamond) embedded in the F ring, designated as "F-ring moonlet". The coefficient of restitution has been set to 0.5 in the radial and longitudinal directions. The solid line designates the location of the F-ring core. Coordinates are in a longitude/radius frame. (a) 0.3 orbits after the impact, (b) 0.45 orbits after the impact, (c) 1.72 orbits after impact (d) 3056 orbits after impact.*

The cloud shears with time and its inclination relative to the F ring orbit decreases with time, as a result of shearing. On long-term evolution (Fig.6d) a spiral is formed and is centred around the F ring orbit as expected (because the massive body's orbit is precisely the same as the F ring's orbit in this configuration). As in the previous case (see section 3) a hole is formed in the central region of the cloud, around the F-ring moonlet: it is the consequence of both (i) the gravitational scattering of particles approaching the satellite and (ii) the accretion of particles onto the satellite itself. This central void, as well as the shape of the cloud of particles shows striking similarities with the jets observed in Cassini images: in Fig.1 a jet is clearly visible at 200° longitude in March and April 2007, whereas the jet's arms see their angle with respect to the F-ring core decrease with time (Murray et al., 2008, Fig.1, Fig.2 of the present paper). A central void is also visible in the same image. Its size is expected to increase with time (as seen in simulations) due to the keplerian shear. We note also that the central void seen in Cassini images corresponds to a void in the F-ring core (Fig.2). This suggests that the full thickness of the F-ring core has been disrupted and scattered away after the impact. A possible implication is that that the core thickness is smaller than, or comparable to, the diameter of a moonlet (about 1km, see Porco et al., 2005), which is in agreement with observations reported by Murray et al. (2008) unveiling a narrow core component small than 1km. So we conclude that on a short timescale, the jet's structure is similar to our simulations, and on long-term evolution, the spiral seen in Cassini images is also compatible with our simulations (Fig. 6d).

## 5. Discussion and conclusion

We have shown that physical collisions between a massive moonlet and massless particles is a mechanism able to scatter material over hundreds of kilometres (see sections 1 and 2), even for a small (km-sized) moonlet. Gravitational scattering by a same-size moonlet would induce a dispersion of only a few kilometres. On the one hand, conservation of momentum during the impact implies that the particles cloud will gather, and spread, around the satellite's orbit (fig. 3.d). On the other hand, Cassini observations (Fig.1, Fig.2) suggest that spirals and jets are all centered around, and parallel to, the F-ring core itself. As a direct consequence (section 4) the best working scenario for reproducing observations is that loose and cohesion less clumps, exterior to the ring collide onto one, or several, moonlets embedded *inside* the F-ring core. On a long term evolution, the spiral arm (made by the ejecta of the clump) will be parallel to the F-ring core, like in observations (Fig.6.d).

Is such a scenario realistic? Recent ISS-Cassini observations have revealed the high complexity of the F-ring core (Murray et al. 2008) with a variety of dynamical structures, some of which could be explained by the presence of massive km-sized moonlets (Murray et al. 2008). Recently UVIS and VIMS tracked some stellar occultations (Nicholson et al. 2007, Esposito et al. 2008) and detected, in the F-ring core, clumps of at least a few hundred meters, like objects designated as "Mittens", "Pywacket", "Butterball" or "Fluffy" in Esposito et al. (2008) with sizes ranging from 600m to 1500m. Such bodies would have only little gravitational influence on their surrounding, whereas, the present paper shows that if they are solid and suffer collisions with other clumps, they would scatter material over hundreds of kilometers. In addition, the presence of bodies of non-negligible mass, inside and the F ring, has been suspected for long (Cuzzi and Burns 1988, Murray et al. 1997). Furthermore, substantial accretion is expected at the location of the F ring (Barbara and Esposito 2002). So there are several good reasons to believe that the F-ring core could shelter a full population of solid km-sized bodies. The F ring is also surrounded by a population of moonlets, or elongated clumps (among which S2004 S6 is the most famous member, see Porco et al., 2005) on orbits crossing the F-ring core (Murray et al., 2005, Spitale et al.2005). When they meet any dense material embedded in the F-ring core, they could be scattered as described in the present paper, and so, have limited lifetime. This may explain why some clumps, detected in the F-ring neighborhood, have disappeared after a few orbits (Murray et al. 2005a). If a portion of the clump still survives after the impact and remains on a similar trajectory, it will re-collide with the F ring at about the same location, thus creating another jet (if a solid body is still present there) next to the previous one, as visible in Fig.1.Quantifying precisely the lifetime of these objects is very difficult because we know too little about the population of bodies inside, and exterior to, the F ring.

What is the source of the clumps? As suggested by different authors (see Esposito et al., 2008 for a review) substantial accretion could be present in the F ring core, because it lies precisely at the Roche Limit for ice. So a "cycle of material" could be present in which clumps are first gravitationally accreted in the F-ring core, then scattered away due gravitational encounters with Prometheus or Pandora or collisions with other clumps, and thus creating jets and spirals structures. Such a scenario, still under debate, may also explain why sudden brightening events in the F ring, on the timescale of weeks, have been observed in Voyager images (Showalter 1998, Poulet et al. 2000). It was hypothesized that they could be the result of meteoroid impacts on the F ring, scattering material over 100 km radially (Showalter 1998). We propose here an alternative explanation: the transient bright spots that Voyager has observed could be the result of a collision of a moonlet with the F ring, as described in the present paper.

In conclusion, it seems that small kilometer-sized moonlets create large-scale structures (of a few tens to a few hundred kilometers) that could be the observed F-ring jets and spirals due to physical collisions with dust and clumps, whereas at small scale (of a few kilometers), they produce delicate gravitational structures as described in Murray et al. 2005b, 2008.

We think that we solve in the present paper the problem raised in Charnoz et al.(2005) that needed Prometheus-sized moons to create jets and spirals. In the vision we propose here, that is basically the same as in Cuzzi & Burns (1988), the F-ring core destroys, feeds, and scatters the surrounding clump (or moonlet) population. Spirals and jets are the direct manifestation of this collisional activity. Both systems are deeply coupled, and the whole forms a formidable "ecosystem" that exchanges material and energy.

**Acknowledgement**: I would like to thank all the Cassini Imaging Team for enlightening discussions. I am indebted to Joe Burns and Carl Murray for their careful reading and comments about this manuscript. I also thank B. Scharringhausen, as a reviewer, whose comments increased the quality of the paper.

# BIBLIOGRAPHY
Barbara J.M., Esposito L.W., 2002. Moonlet collisions and the effect of tidally modified accretion in Saturn's F ring. *Icarus* **160**, 161-171

Bosh, A. S.; Olkin, C. B.; French, R. G.; Nicholson, P. D. 2002. Saturn's F Ring: Kinematics and Particle Sizes from Stellar Occultation Studies. *Icarus* **157**, 57-75

Bridges F.G., Hatzes A., Lin D.N.C., 1984. Structure, stability and evolution of Saturn's rings. *Nature* **309**, 333-335

Charnoz, S.; Thébault, P.; Brahic, A, 2000.Short-term collisional evolution of a disc perturbed by a giant-planet embryo. *A&A* **373**, 683-701

Charnoz, S., Porco C. C., Déau E., Brahic A., Spitale J. N., Bacques G., Baillie K. 2005. Cassini Discovers a Kinematic Spiral Ring Around Saturn. *Science* **310**, 1300-1304

Cuzzi, J. N.; Burns, J. A. 1988. Charged particle depletion surrounding Saturn's F ring - Evidence for a moonlet belt? *Icarus* **74**, 284-324.

Esposito L.W., Meinke B.K., Colwell J.E., Nicholson P.D., Hedman M.H., 2008. Moonlets and clumps in Saturn's F ring. *Icarus* **194**, 278-289

Smith B.A., and 26 co-authors, 1981. Encounter with Saturn: Voyager 1 imaging science results. *Science* **212**, 163-191.

Murray C.D., Gordon M.K., Giuliatti Winter S.M., 1997. Unraveling the strands of Saturn's F Ring. *Icarus* **129**, 304-316

Murray C.D., Dermott S.F., 1999. Solar System Dynamics (Cambridge University Press)

Murray C.D., Evans M.W., Cooper N., Beurle K., Burns J.A., Spitale J., Porco C.C., 2005a. Saturn's F ring and its retinue. *BAAS* **37**, p767

Murray C.D., Chavez C., Beurle K., Cooper N., Evans M.W., Burns J.A., Porco C.C., 2005b. How Prometheus creates structure in Saturn's F ring. *Nature* **437**, 1326-1329

Murray C.D., Beurle K., Cooper N.J., Evans M.W., Williams G.A., Charnoz S., 2008. The determination of the structure of Saturn's F ring by nearby moonlets. To appear in *Nature*

Nicholson P.D., Hedman M.M., Wallis B.D., and the Cassini VIMS team, 2007. Cassini-VIMS observations of stellar occultations by Saturn's rings. DDA meeting #38, #12.05

Porco C.C., Baker E., Barbara J., Beurle K., Brahic A., Burns J.A., Charnoz S., Cooper N., Dawson D.D., Del Genio A.D., Denk T., Dones L., Dyudina U., Evans M.W., Giese B., Grazier K., Helfenstein P., Ingersoll A.P., Jacobson R.A., Johnson T.V., McExenA., Murray C.D., Neujum G., Owen W.M., Perry J., Roatsch T., Spitale J., Squyres S., Thomas P., Tiscareno M., Turtle E., Vasavada A.R., Veverka J., Wagner R., West R., 2005. Cassini imaging science : initial results on Saturn's rings and small satellites. *Science* **307**, 1226-1236

Poulet F., Sicardy B., Nicholson P.D., Karkoschka E., Caldwell J., 2000. Saturn's Ring-Plane Crossings of August and November 1995: A Model for the New F-Ring Objects. *Icarus* **144**, 135-148

Richardson, D. C. 1994. Tree Code Simulations of Planetary Rings. *MNRAS* **269**, 493- 511

Showalter M.R., Burns J.A., 1982. A numerical study of Saturn's F ring. *Icarus* **52**, 526-544

Showalter M.R., Pollack J.B., Ockert M.E., Doyle L.R., Dalton J.B., 1992. A photometric study of Saturn's F ring. *Icarus* **100**, 394-411

Showalter M.R., 1998. Detection of Centimeter-Sized Meteoroid Impact Events in Saturn's F Ring. *Science* **282**, 1099-1102

Showalter, M. R. 2004. Disentangling Saturn's F Ring. I. Clump orbits and lifetimes. *Icarus* **171**, 56-371

Spitale J.N., Jacobson R.A., Porco C.C., Owen W.M., Charnoz S., 2005. The orbits of Saturn's small satellites. *BAAS* **37**, p529

Supulver K.D., Brigdges F.G., Lin D.N.C., 1995. The coefficient of restitution of ice particles in glancing collisions: experimental results for unfrosted surfaces. *Icarus* **113**, 188-199

# TABLE

| $\varepsilon_n$, $\varepsilon_t$ | 0.6 | 0.1 | 0.06 | 0.02 |
|---|---|---|---|---|
| Radial spreading around the moonlet's orbit | ±1300 km | ±200 km | ±80 km | ±30 km |

**Table 1**: Radial spreading of material scattered by the moonlet as a function of the coefficient of restitution observed in numerical simulations.